\documentclass[prd,amsmath,aps,floats,amssymb, floatfix,
  superscriptaddress,nofootinbib,twocolumn,preprintnumbers]{revtex4-2}
\usepackage{tabularx}
\usepackage{amsmath,amssymb,natbib,latexsym,times}

\usepackage{booktabs}
\usepackage{multirow}

\usepackage{graphicx}% Include figure files
\usepackage{dcolumn}% Align table columns on decimal point
\usepackage{bm}% bold math

\usepackage{url}
\usepackage{color}
\usepackage{comment}
\usepackage{mathrsfs}
\usepackage{hyperref}% add hypertext capabilities
\usepackage[dvipsnames]{xcolor}
\usepackage{orcidlink}
\usepackage{mathtools}

\newcommand{\mnras}{Monthly Notices of the Royal Astronomical Society}
\newcommand{\physrep}{Physics Reports}
\newcommand{\jcap}{Journal of Cosmology and Astroparticle Physics}

\renewcommand{\apj}{The Astrophysical Journal}

\newcommand{\aap}{Astronomy and Astrophysics}

\newcommand{\Om}{\Omega_\mathrm{m}}

\begin{document}

\title[]{
Late-time suppression of structure growth as a solution for the $S_8$ tension}

\author{Ryo~Terasawa\orcidlink{0000-0002-1193-623X}}
\email{ryo.terasawa@ipmu.jp}
\affiliation{Kavli Institute for the Physics and Mathematics of the Universe (WPI), The University of Tokyo Institutes for Advanced Study (UTIAS), The University of Tokyo, Chiba 277-8583, Japan}
\affiliation{Department of Physics, The University of Tokyo, Bunkyo, Tokyo 113-0031, Japan}
\affiliation{%
Center for Data-Driven Discovery (CD3), Kavli IPMU (WPI), UTIAS, The University of Tokyo, Kashiwa, Chiba 277-8583, Japan}%

\author{Masahiro~Takada\orcidlink{0000-0002-5578-6472}}
%\email{masahiro.takada@ipmu.jp}
\affiliation{Kavli Institute for the Physics and Mathematics of the Universe (WPI), The University of Tokyo Institutes for Advanced Study (UTIAS), The University of Tokyo, Chiba 277-8583, Japan}
\affiliation{%
Center for Data-Driven Discovery (CD3), Kavli IPMU (WPI), UTIAS, The University of Tokyo, Kashiwa, Chiba 277-8583, Japan}%

\author{Toshiki~Kurita\orcidlink{0000-0002-1259-8914}}
\affiliation{Max Planck Institute f\"ur Astrophysik, Karl-Schwarzschild-Str. 1, 85748 Garching, Germany}
\affiliation{Kavli Institute for the Physics and Mathematics of the Universe (WPI), The University of Tokyo Institutes for Advanced Study (UTIAS), The University of Tokyo, Chiba 277-8583, Japan}

\author{Sunao~Sugiyama\orcidlink{0000-0003-1153-6735}}
\affiliation{Center for Particle Cosmology, Department of Physics and Astronomy, University of Pennsylvania, Philadelphia, PA 19104, USA}
\affiliation{Kavli Institute for the Physics and Mathematics of the Universe (WPI), The University of Tokyo Institutes for Advanced Study (UTIAS), The University of Tokyo, Chiba 277-8583, Japan}

\begin{abstract}
{The $S_8$ value inferred from 
the Subaru Hyper Suprime-Cam (HSC) Year~3 cosmic shear data, under the assumption of the 
flat $\Lambda$CDM model, is 2--3$\sigma$ lower than that 
inferred from observations of the early-time universe, such as cosmic microwave background (CMB) anisotropy data.
Resolving the $S_8$ tension requires a scenario in which structure formation on small scales is suppressed in the late universe. 
As potential solutions,
we consider extended models both within and beyond the $\Lambda$CDM model -- models that 
incorporate
parameterized baryonic feedback effects, the effect of varying neutrino mass, and modified structure growth, each of which can lead to a suppression of structure growth at lower redshifts, with its own distinct scale- and redshift-dependencies. 
In particular, we consider phenomenological modified gravity models in which the suppression of structure growth is triggered at lower redshifts, as dark energy ($\Lambda$)
begins to dominate the background expansion. 
We show that the modified growth factor models -- especially 
those featuring more rapid growth suppression at lower redshifts -- 
provide an improved fit to the combined datasets of 
the HSC-Y3 cosmic shear correlation functions, the {\it Planck} CMB, and the ACT DR6 CMB lensing, compared to
the fiducial $\Lambda$CDM model and 
the models including the baryonic effects or the massive neutrino effect within the 
the $\Lambda$CDM framework.}
\end{abstract} 

\maketitle

\section{Introduction}
\label{sec:intro}
The standard model of the universe, the flat $\Lambda$ Cold Dark Matter ($\Lambda$CDM) model, has been successfully explaining a variety of observations \citep[e.g.,][]{2013PhR...530...87W,2020moco.book.....D}. 
{The fundamental assumptions underlying the $\Lambda$CDM model include the initial conditions predicted by inflation, General Relativity (GR) as the theory of gravity, the existence of cold dark matter, and the dominance of the cosmological constant ($\Lambda$), which drives the accelerated expansion of the late-time universe.}

Despite the success of the $\Lambda$CDM model, recent high-precision observations 
have {suggested} {potential} discrepancies with the standard model.
One such discrepancy is known as the $\sigma_8$ or $S_8$ tension, {where $\sigma_8$ (or $S_8$) characterizes the clustering amplitude in the present-day universe.}
This refers to the consistent lower values of $\sigma_8$ or $S_8$ in $\Lambda$CDM models inferred from large-scale structure (LSS) probes, 
compared to those inferred from the \textit{Planck}-2018 cosmic microwave background (CMB) measurements~\citep[see][for a recent review]{2022JHEAp..34...49A}.
Such large-scale structure probes that exhibit the $S_8$ (or $\sigma_8$)
tension include 
cosmological weak lensing \citep[hereafter ``cosmic shear'', e.g.,][]{HSC3_cosmicShearReal,HSC3_cosmicShearFourier,KiDS1000_CS_Asgari2020,DESY3_CS_Secco2022},
joint probe cosmology combining weak lensing and galaxy clustering
\citep[e.g.,][]{HSC3_3x2pt_ss,HSC3_3x2pt_ls,KiDS1000_3x2pt_Heymans2021,DESY3_3x2pt2022}
and redshift-space galaxy clustering \citep[e.g.,][]{2020JCAP...05..042I,2022JCAP...02..008C,2022PhRvD.105h3517K}. 
This tension provides an opportunity to explore {extensions of the model}
by relaxing the {assumptions of} $\Lambda$CDM.
For example, in \citet{2025arXiv250320396T}, {we relaxed the inflationary assumption of a single 
power-law form for the primordial power spectrum and explored the possibility of alleviating the tension by modifying the primordial power spectrum.}

Another possibility for solving the tension is to relax the assumption of the GR or the cosmological constant $\Lambda$,
 by introducing a modified gravity theory or a dark energy model~\citep{NHW_gamma, MX24,2024arXiv241009499A}. 
{An effective and widely used approach to exploring a model beyond the $\Lambda$CDM framework
is to modify the linear growth rate of structure formation and test whether the modified model can consistently explain cosmological datasets from both the early and late universe.
One such model is the $\gamma$ growth index model \citep{2005PhRvD..72d3529L} in which 
the linear growth rate is empirically modeled as $f=\Omega_{\rm m}(z)^\gamma$, $\gamma\simeq 0.55$ corresponds to the GR prediction. Ref.~\citep{2017MNRAS.465.1757G} compared the $\gamma$ index model with 
the SDSS BOSS data and found  $\gamma\sim 0.7$, which is about 2.7$\sigma$ away from the GR prediction.}
A larger value of $\gamma$ than 0.55 leads to suppressed structure growth in the late-time Universe, which helps mitigate the $S_8$ tension.

{An extension within the $\Lambda$CDM model might also mitigate the tension.
Such variants include baryonic effects arising from the physics of galaxy formation and evolution
\citep[e.g.][]{Chisari18_Horizon-AGN,Jiachuan23, Terasawa24}, 
modifications to the nature of cold dark matter (e.g., ultralight axions)~\citep{2025arXiv250206687P}, and the inclusion of massive neutrino effects, where the neutrino mass is treated as a free parameter in the inference.
These extensions lead to suppressed structure growth on small scales.}

In this paper, {we examine whether extended models, either within or beyond the $\Lambda$CDM framework, can consistently explain various cosmological datasets from both the early and late Universe -- specifically, CMB anisotropy data, CMB lensing, and cosmic shear data.
For the cosmic shear data, we use the measurements from the Subaru Hyper Suprime-Cam Year~3 (HSC-Y3) dataset~\cite{wlreana_catalogs_Longley2023,HSC3_cosmicShearReal,Terasawa24}, which indicates the $S_8$ tension.
For the extended models, we consider those that incorporate baryonic effects, the effects of massive neutrinos, and phenomenological models inspired by modified gravity theories, where the modifications are parameterized by additional parameters. In particular, we use modified gravity models in which modifications to the linear growth factor are triggered at redshifts where dark energy ($\Lambda$) dominates the expansion history.}

We organize this paper as follows. In Section~\ref{sec:models}, we introduce models for the structure growth suppression and discuss the effects on the cosmic shear signal.
In Section~\ref{sec:analysis}, we explain the details of the joint analysis of the cosmological datasets presented in this paper, and we will show the results in Section~\ref{sec:results}.
Section~\ref{sec:conclusion} is devoted to conclusion.

\section{Suppression of Structure Formation in Extended Models Within and Beyond $\Lambda$CDM}
\label{sec:models}
{In this section, we describe extended models used in this paper that lead to a 
suppression of structure formation in the late universe, motivated by the possibility that such models might 
alleviate the $S_8$ tension. We first describe extended models within the $\Lambda$CDM framework,
and then describe those inspired by modified gravity beyond $\Lambda$CDM.}\\

\begin{figure}
\includegraphics[width=\columnwidth]{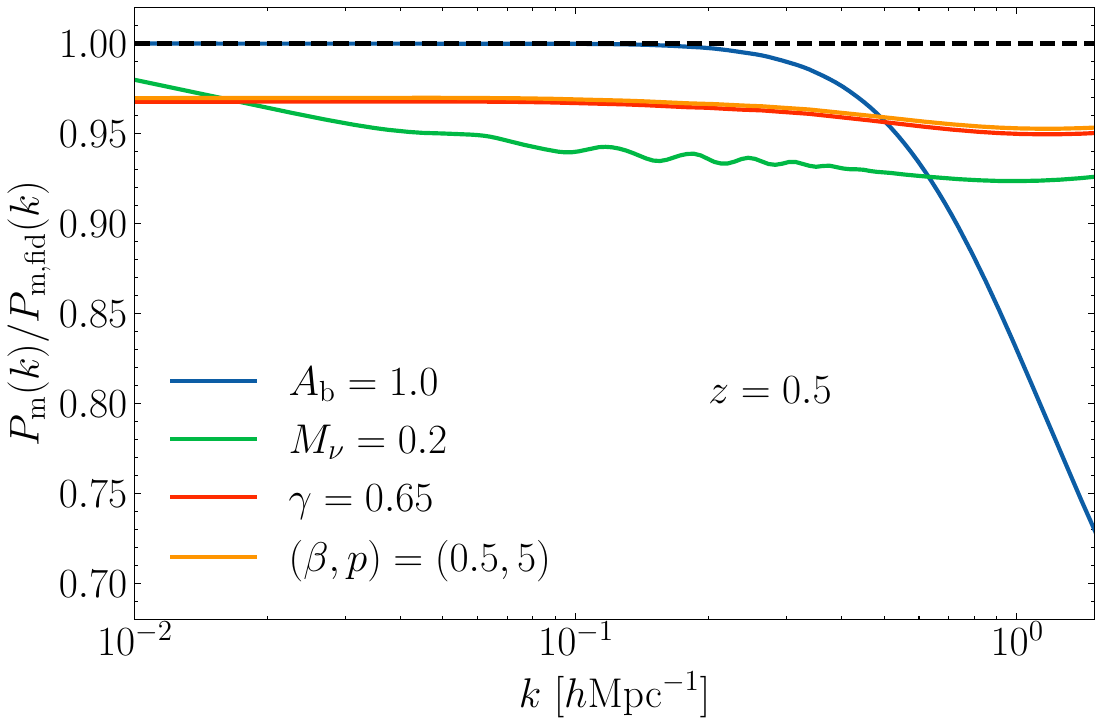} 
    \caption{
    Fractional change in the {nonlinear} matter power spectrum $P_{\rm m}(k)$ at $z=0.5$ 
    due to the baryonic effects, {the massive neutrino effect, and the modified structure growth, 
    relative to the fiducial $\Lambda$CDM model}. {Here we consider 
    the baryonic effect model parametrized by $A_{\rm b}$ (Section~\ref{subsubsec:HMCode16}), the massive neutrino model characterized by the total neutrino mass $M_\nu$ (Section~\ref{subsec:massive_neutrinos}), and the modified growth model parameterized 
    by $\gamma$ (Section~\ref{ssub:growth_index}) {or $(\beta, p)$ (Section~\ref{ssub:detg})}. 
    The fiducial $\Lambda$CDM model corresponds to $(A_{\rm b},\gamma, \beta, M_{\nu})=(3.13,0.55, 0.0,0.06~{\rm eV})$.
    Note that, throughout this paper, we use the CMB normalization for the linear power spectrum.
    We vary each of the parameters as given in the legend to compute the nonlinear matter power spectrum 
    in the numerator on the $y$ axis. 
    When varying $M_{\nu}$, we keep $\Omega_{\rm c}h^2$, $\Omega_{\rm b}h^2$ and $\Omega_{\rm m}$ fixed
    to ensure that the CMB spectra in the early universe remain nearly unchanged.
    All the extended models with these parameter values lead to a suppression of the power spectrum amplitudes at low redshifts relevant to cosmic shear.}
    }
    \label{fig:Pkratio}
\end{figure}

\subsection{{Extended models within the $\Lambda$CDM model}}
\label{subsec:scale-dependent}

{Even within the $\Lambda$CDM framework, there are effects that can lead to a suppression of structure formation on small scales. Representative examples include the baryonic effects associated with the physics of galaxy formation and evolution \citep{2011MNRAS.417.2020S, Chisari18_Horizon-AGN,2019OJAp....2E...4C,halofit_mead21,Amon_Efstathiou22,baryon_arico23,Terasawa24}, as well as the effects of massive neutrinos \citep{1980PhRvL..45.1980B,1998PhRvL..80.5255H,2006PhRvD..73h3520T}.}

\subsubsection{\textsc{HMCode16} baryonic feedback model}
\label{subsubsec:HMCode16}

{Since galaxy formation involves complex physical processes, we adopt simplified models commonly used in weak lensing cosmology, in which baryonic effects are characterized by a small set of parameters calibrated against cosmological hydrodynamical simulations.
In this paper, we consider two models for the baryonic effects, 
following the method in \citet{Terasawa24}.}

{The first model we use is}
\textsc{HMCode16}~\citep{halofit_mead16}, which {models}
the effects of baryonic feedback {using the}
halo bloating parameter $\eta_\mathrm{b}$ and the amplitude of the halo
mass-concentration relation $A_\mathrm{b}$~\citep{halofit_mead15,
halofit_mead16}. 
We follow \citet{Joachimi2020} to set the bloating parameter as a
function of the amplitude parameter:
\begin{equation}
    \eta_\mathrm{b} = 0.98 - 0.12A_\mathrm{b}\,.
\end{equation}
{In \textsc{HMCode16}},
 $A_{\rm b}=3.13$ corresponds to the
matter power spectrum without the baryonic feedback effect (i.e. the spectrum obtained
from dark matter only simulations).
This {relatively} simple parametrization 
{has been}
shown to be sufficient {for obtaining an}
unbiased measurement of $S_8$ from the HSC-Y3 cosmic shear 2PCFs, even in the presence of the baryonic feedback, {if the appropriate scale cuts are employed~\citep{HSC3_cosmicShearReal}.}
We consider {a} more complicated model in Section~\ref{subsubsec:HMCode20}. 

In this paper, we adopt the flat prior on $A_{\rm b}$ given as $A_{\rm b} = [1.0, 4.0]$. 
In Fig.~\ref{fig:Pkratio}, we show the effect of $A_{\rm b}=1$ on the matter power spectrum compared to the dark matter only nonlinear power spectrum.
The lower bound $A_{\rm b} = 1$ roughly corresponds to the {effect} seen in the
COSMO-OWLS simulation~\cite{cowls_LeBrun2014} assuming
the strongest AGN feedback, $T_{\rm AGN} = 10^{8.7}~K$ (see e.g., Fig.2 of \citet{Terasawa24}), and {is} stronger than {the effect estimated from} 
the combined analysis of the cosmic shear and kinetic SZ effect in \citet{Bigwood24}. 
{The figure shows that the baryonic effect does not alter the matter power spectrum on large scales but leads to significant suppression on small scales due to its causal nature{; the influence of baryonic physics originates on small scales within a halo and subsequently propagates to larger scales at a finite speed.}

\subsubsection{\textsc{HMCode20} baryonic feedback model}
\label{subsubsec:HMCode20} 
We also {use \textsc{HMCode20}~\citep{halofit_mead21}, a}
more flexible {model} for the baryonic effect than \textsc{HMCode16}.
The model describes {the} clustering of three components based on the halo model: dark matter, stellar component that approximates a central galaxy by a point mass in its host halo, and diffuse intra-halo gas. Then \textsc{HMCode20} models relative fractions of the central galaxy and the diffuse gas in the total budget of baryonic matter, and models variations in the radial profile of total matter (dark matter plus the diffuse gas) in halos by changes in the halo concentration, from the DM-only Navarro-Frenk-White profile~\citep{NFW}, as a function of halo mass. \textsc{HMCode20} uses six parameters as given in Table~\ref{tab:parameters}: the stellar mass fraction ($f_{\ast,0}$); the characteristic halo mass scale $M_{b,0}$ below which diffuse gas is blown away from the host halo by the baryonic feedback effect and above which intra-halo diffuse gas is confined in the gravitational potential well of the host halo; and the normalization parameter $B_0$ for the halo concentration. And it has the three additional parameters to describe the redshift dependence of the three parameters. Changing these parameters can reproduce the DM-only model and can
describe both enhancement and suppression in the matter power spectrum amplitudes, which physically correspond to the baryon contraction \citep[e.g.][]{2004ApJ...616...16G} and the baryonic feedback effects of supernovae and AGNs \citep[e.g.,][]{2003ApJ...599...38B,2015JCAP...12..049S,baronCorrect_Schneider2019}, respectively.

\citet{halofit_mead21}
calibrated the model parameters against
\textsc{COSMO-OWLS} \cite{cowls_LeBrun2014}
and \textsc{BAHAMAS} \cite{2018MNRAS.476.2999M} hydrodynamical simulation suites. 
The model is flexible enough to fit the other hydrodynamical simulations that are not used for the calibration, including \textsc{Horizon-AGN} \citep{Chisari18_Horizon-AGN}, \textsc{SIMBA} \citep{2019MNRAS.486.2827D},
    \textsc{Illustris} \citep{Illustris_Vogelsberger2014}, \textsc{Illustris-TNG300} \citep{2018MNRAS.475..676S}, and \textsc{Eagle}~\citep{eagle_Schaye15}, at $\sim 5\%$ level of accuracy~\citep{2025MNRAS.537.1749M}.

Furthermore, \citet{halofit_mead21} proposed an effective model of the baryonic effect given by a single parameter, $\Theta_{\rm AGN}=\log_{10}(T_{\rm AGN}/{\rm K})$, motivated by the fact that the AGN feedback is the most important on the scales relevant to the cosmic shear signal and $T_{\rm AGN}$ roughly corresponds to the heating temperature due to the AGN feedback, which is responsible for the expansion effect of gas from the host halos. \citet{halofit_mead21} showed that 
variations in the six parameters found from a calibration set of the hydrodynamical simulations are fairly well captured by changes in $T_{\rm AGN}$ \citep[see Table~5 of][]{halofit_mead21}.

\citet{Terasawa24} showed that the HSC-Y3 cosmic shear 2PCFs can be {reconciled}
with the {\it Planck} cosmology {\it if} we allow $\Theta_{\rm AGN} \simeq 9.8$. Hence, we impose the priors on the six baryonic feedback parameters, {restricting the feedback strength to correspond to}
$\Theta_{\rm AGN} \leq 10.0$.
Note that $\Theta_{\rm AGN} \gtrsim 9$ {corresponds to stronger}
feedback than that found in any hydrodynamical simulation to date and is considered an unrealistic representation of baryonic effects.
{Nevertheless, we adopt such broad priors to investigate the potential impact of the largest possible baryonic effects on our results.}

\subsubsection{Massive neutrinos}
\label{subsec:massive_neutrinos}

Massive neutrinos suppress the matter power spectrum at scales smaller than its free-streaming scale \cite{1980PhRvL..45.1980B,2008PhRvL.100s1301S}.
{For the fiducial $\Lambda$CDM model, we assume $M_\nu=0.06$~eV for the total neutrino mass,  
corresponding to the lower bound of the normal mass hierarchy \citep[e.g.,][]{2006PhRvD..73h3520T}.
However, the exact value of total neutrino mass is not yet known, and the true value might differ from the assumed value, $0.06$~eV. Therefore, we consider models with varying neutrino mass in the cosmological inference.}

{We first use \textsc{CAMB}\citep{CAMB2000} to
compute the linear matter power spectrum for a given value of the total neutrino mass at a given redshift. We then use}
\textsc{HMCode16}~\citep{halofit_mead16} to model the effect of the massive neutrinos on
the nonlinear matter power spectrum, based on the mapping 
from the input linear power spectrum.

In Fig.~\ref{fig:Pkratio}, we show the effect of the massive neutrinos with the total mass of $M_{\nu} = 0.2~$eV, compared to the fiducial $\Lambda$CDM model with 
$M_{\nu} = 0.06~$eV. Note that, when we {change
the neutrino mass, we keep 
the physical density parameter of CDM 
($\Omega_{\rm c}h^2$) and baryon ($\Omega_{\rm b}h^2$), and 
the present-day density parameter of matter ($\Omega_{\rm m}$) 
fixed within the flat $\Lambda$CDM, which means that the Hubble parameter $h$ needs to be changed accordingly. 
Doing so keeps the CMB spectra nearly unchanged.}
{A change in neutrino mass alters the matter power spectrum at low redshifts relevant to cosmic shear over a wide range of wavenumbers, not only on small nonlinear scales, but also on linear scales.
The constraining power of cosmic shear primarily comes from nonlinear scales, but the scale dependence on these scales, around $k\sim 1~h{\rm Mpc}^{-1}$, is weak.
In addition, the redshift dependence of the neutrino effect within the redshift range relevant to cosmic shear is weak.
The scale and redshift dependencies of the neutrino effect are different from those of the baryonic effect, which help to distinguish between these two effects using cosmological data.}

\subsection{Extended Models beyond $\Lambda$CDM: Modified Gravity}
\label{subsec:scale-independent}

\begin{figure}
\includegraphics[width=\columnwidth]{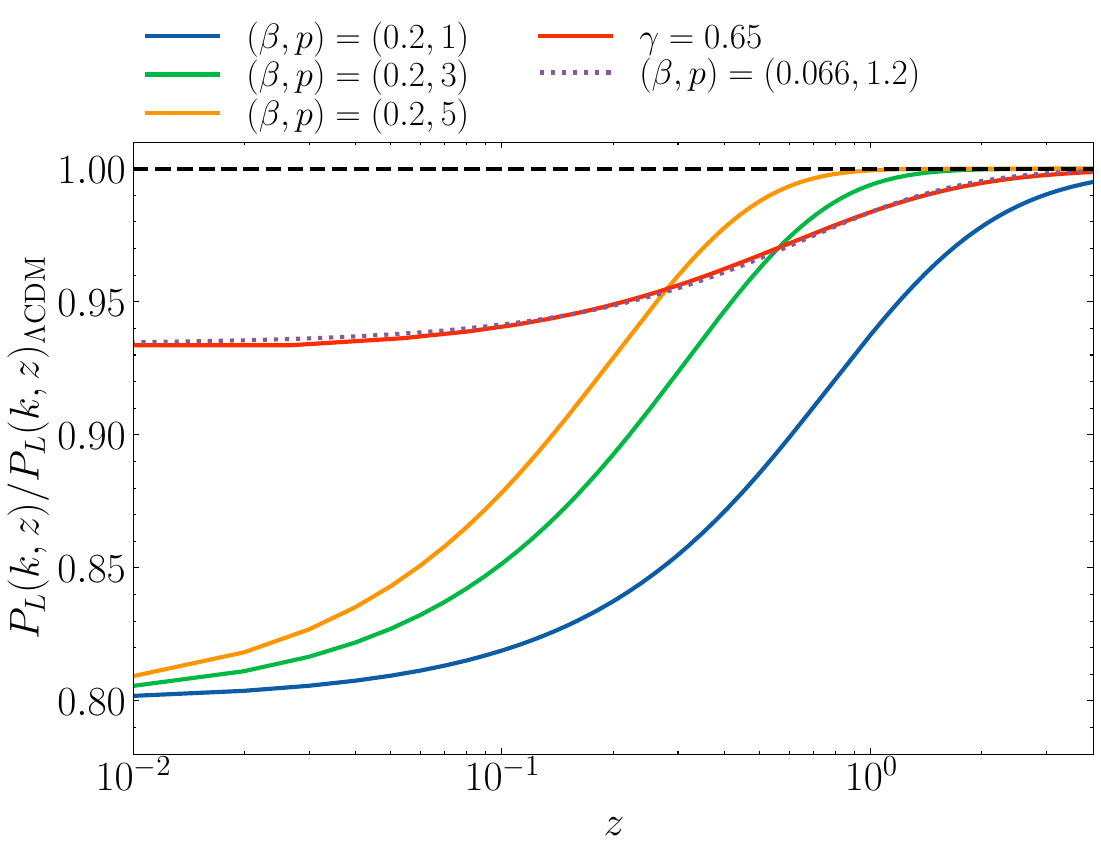} 
    \caption{
    Fractional change in the linear matter power spectrum $P_L(k,z)$ 
    due to the {modified growth factor models, relative to the fiducial 
    $\Lambda$CDM model, plotted as a function of redshift. 
    Here, we consider the $\gamma$ growth index model as in Fig.~\ref{fig:Pkratio}, and 
    the dark energy tracking growth (DETG) model, parameterized by two parameters, 
    $\beta$ and $p$.
    The DETG model with $\beta=0$ reproduces the $\Lambda$CDM model by construction.
    We chose the parameter values leading to a suppression of structure growth compared to $\Lambda$CDM model. Both models predict a scale-independent change in the amplitude of the linear power spectrum.
    Different values of the model parameters give different redshift dependencies of the growth suppression.
    We also demonstrate that the DETG model with $\beta=0.066$ and $p=1.2$ exhibits
    a redshift dependence similar  to that of the 
    $\gamma$ model with $\gamma=0.65$.}
    }
    \label{fig:Pzratio}
\end{figure}

{Another class of scenarios that lead to a suppression of structure formation is based on modified gravity models. This corresponds to extended models beyond the $\Lambda$CDM framework.
Once the possibility of modified gravity is allowed, there is an infinitely broad class of models to consider. In this paper, we consider two phenomenological models for modified gravity in the context of structure formation, both commonly used in weak lensing cosmology.}

\subsubsection{$\gamma$ growth index model}
\label{ssub:growth_index}

{The first modified model is the $\gamma$ growth index model, in which the linear growth factor of structure formation is modified and the modification is simply parameterized by a single parameter, $\gamma$ 
\cite{2005PhRvD..72d3529L}. 
In this model,}
The linear growth rate $f(a) \equiv \mathrm{d}\ln D(a)/\mathrm{d}\ln a$, {where} $D(a)$ is the linear growth factor as a function of the scale factor $a$, is modified as
\begin{align}
    f(a) = \Omega_{\rm m}(a)^{\gamma},
\end{align}
where $\gamma$ is the growth index {parameter} and $\Omega_{\rm m}(a)$ is the fractional energy density of matter at a given epoch corresponding to the scale factor 
$a$. 
Note that, for the $\gamma$ model,  
we keep the background expansion history the same as that of the fiducial
$\Lambda$CDM model; therefore $\Omega_{\rm m}(a)$ in the above equation is given by the background model.
$\gamma\simeq 0.55$ corresponds to the standard $\Lambda$CDM model {based on the assumption of General Relativity.
We then obtain the linear growth factor by integrating the above equation:
\begin{align}
    D(a; \gamma) 
    {\propto \exp \left[\int^a \!\mathrm{d}a' \frac{\Omega_{\rm m}(a')^{\gamma}}{a'}\right].}
    \label{eq:D_gamma}
\end{align}
{Throughout this paper we adopt the CMB normalization, specified by $A_s$ (see below), 
for the linear power spectrum.}
Therefore, we normalize $D(a)$ such that
$D(a)/a\propto 1$ in the matter dominated regime such as the CMB epoch at 
{$z\simeq1100$}.
}
{With this modified growth rate,}
we calculate the linear power spectrum as
{
\begin{align}
      P_L(k, a;\gamma) = P_L(k,a)_{\Lambda{\rm CDM}} 
      %D^2(\gamma,a) \\ \nonumber
      \times \left(\frac{D(a;\gamma)}{D(a;\gamma_{\Lambda{\rm CDM}})}
      \right)^2,
\end{align}
where $P_L(k,a)_{\Lambda{\rm CDM}}$ is the linear power spectrum for the $\Lambda$CDM model
and $\gamma_{\Lambda{\rm CDM}}=0.55$.}

{If we allow $\gamma$ to vary and treat it as a free parameter in the parameter inference, it 
provides a phenomenological model of modified gravity.}
Therefore, a deviation of $\gamma$ from $0.55$ indicates a signature of modified gravity
in cosmological data~\cite{2005PhRvD..72d3529L}.
{The $\gamma $ model with $\gamma>0.55$ leads to a suppression of structure growth at lower redshifts, where dark energy becomes significant to the Hubble expansion, compared to the prediction of the $\Lambda$CDM model.}
To compute the nonlinear matter power spectrum with varying $\gamma$, we utilize the modified version of CAMB~\citep{CAMB2000} developed in Refs.~\cite{NHW_gamma, 2023JCAP...09..028W}.

The red line in Fig.~\ref{fig:Pkratio} shows
{how varying $\gamma$ from $0.55$ to $0.65$ changes the nonlinear matter power spectrum.
On linear, large scales, this results in a scale-independent suppression of the power spectrum amplitude.
On the other hand, on nonlinear, small scales, the change exhibits a weak scale-dependence in the 
suppression,
due to the nonlinear mapping from the linear to nonlinear power spectrum~\cite{2022PhRvD.106h3504T}.}
As we {will show below,}
varying the growth index affects the cosmic shear {correlation functions}
in {an} almost scale-independent way over the scales of interest.
{For comparison, the red line in}
Fig.~\ref{fig:Pzratio} illustrates {how varying $\gamma$ alters the amplitude of the linear power spectrum, as a function of redshift on the $x$ axis.

\subsubsection{{Dark energy tracking growth (DETG) model}}
\label{ssub:detg}

\citet{MX24} proposed a phenomenological model that rescales the structure growth as a function of redshift. In this model, the linear matter power spectrum is modified {as a function of redshift} as follows:
\begin{align}
    \alpha(z) &\equiv \frac{P_{L}(k,z)}{P_{L}(k,z)_{\Lambda{\rm CDM}}} \\
    &= \left(\frac{D_{\rm DETG}(z)}{D_{\Lambda{\rm CDM}}(z)}\right)^2
    =  1 - \beta \left(\frac{\Omega_{\rm DE}(z)}{\Omega_{\rm DE}(z=0)} \right)^{p},
    \label{eq:DETG}
\end{align}
where $\Omega_{\rm DE}(z)(=1-\Omega_{\rm m}(z))$ 
is the fractional energy density of dark energy at redshift $z$.
{Here, we assumed that the linear power spectrum remains unchanged in the matter-dominated regime
with $\Omega_{\rm DE}(z)\simeq0$, such as the CMB epoch.} 
{This model is given as a function of two parameters,}
$\beta$ and $p$. {As in the $\gamma$ model, $\Omega_{\rm DE}(z)$ is given by the
background $\Lambda$CDM model. 
Since the modification in structure growth, relative to the GR prediction, is triggered by dark energy domination, characterized by $\Omega_{\rm DE}(z)$,
we refer to this model as the ``Dark Energy Tracking Growth (DETG) model.''}

In Fig.~\ref{fig:Pzratio}, we illustrate {the changes in structure growth, or equivalently, the changes in the linear power spectrum amplitude, for the DETG model.
A positive or negative $\beta$ predicts a suppressed or enhanced amplitude 
in the linear power spectrum at $z=0$, respectively, compared to the $\Lambda$CDM model.
The parameter $p$ determines how rapidly the modification of structure growth evolves with 
redshift in conjunction with the dark energy density.}
{In this paper, we will consider a specific value of $p$ chosen from}
$p =1, 2, 3, 4, 5$ or 6.
We implement the model by modifying \textsc{CAMB}~\citep{CAMB2000}.

We use \textsc{HMCode16}~\citep{halofit_mead16} to model the nonlinear matter power spectrum $P_{\rm m}(k)$ {for the DETG model, using the mapping
from an input linear power spectrum.
Fig.~\ref{fig:Pkratio} shows the prediction for the DETG model with $\beta=0.5$ and $p=5$, which has
a rapid redshift evolution with the onset of dark energy domination. 
{The effect is almost indistinguishable from that of $\gamma=0.65$ at this particular redshift $z=0.5$, although the changes in structure growth predicted by the $\gamma$
model and the DETG model differ at other redshifts as demonstrated 
in Fig.~\ref{fig:Pzratio}.}}

\subsubsection{Relation between $\gamma$ and DETG models}

{We discussed phenomenological models for the modification of structure growth in Sections~\ref{ssub:growth_index} and \ref{ssub:detg}.
Here we discuss a relation between these two models.}

{When $\beta=0$ in Eq.~(\ref{eq:DETG}), the DETG model reproduces the $\Lambda$CDM model under GR.
If we set $D_\gamma=D_{\rm DETG}$ in Eqs.~(\ref{eq:D_gamma}) and (\ref{eq:DETG}),
we can find the relation between the $\gamma$ and DETG models as}
\begin{align}
    \Om(a)^{\gamma} = \Om(a)^{\gamma_{\Lambda{\rm CDM}}} + \frac{1}{2} \frac{\mathrm{d}\ln \alpha(a)}{\mathrm{d}\ln a},
\end{align}
{where we assumed {$\ln D(\gamma_{\Lambda{\rm CDM}},z)/\ln a \simeq \Omega_{\rm m}(z)^{\gamma_{\Lambda{\rm CDM}}}$} with $\gamma_{\Lambda{\rm CDM}}=
0.55$.}
For a given $\gamma$, we can find the effective
parameters $(\beta, p)$ that 
reproduce the {growth factor for the $\gamma$ model.
This is demonstrated in Fig.~\ref{fig:Pzratio}. 
The DETG model with $\beta=0.066$ and $p=1.2$ closely reproduces the 
$\gamma$ model with $\gamma=0.65$.}

\subsection{Effects on Cosmic Shear Correlation Functions}
\label{subsec:cosmic shear}

\begin{figure*}
\includegraphics[width=2\columnwidth]{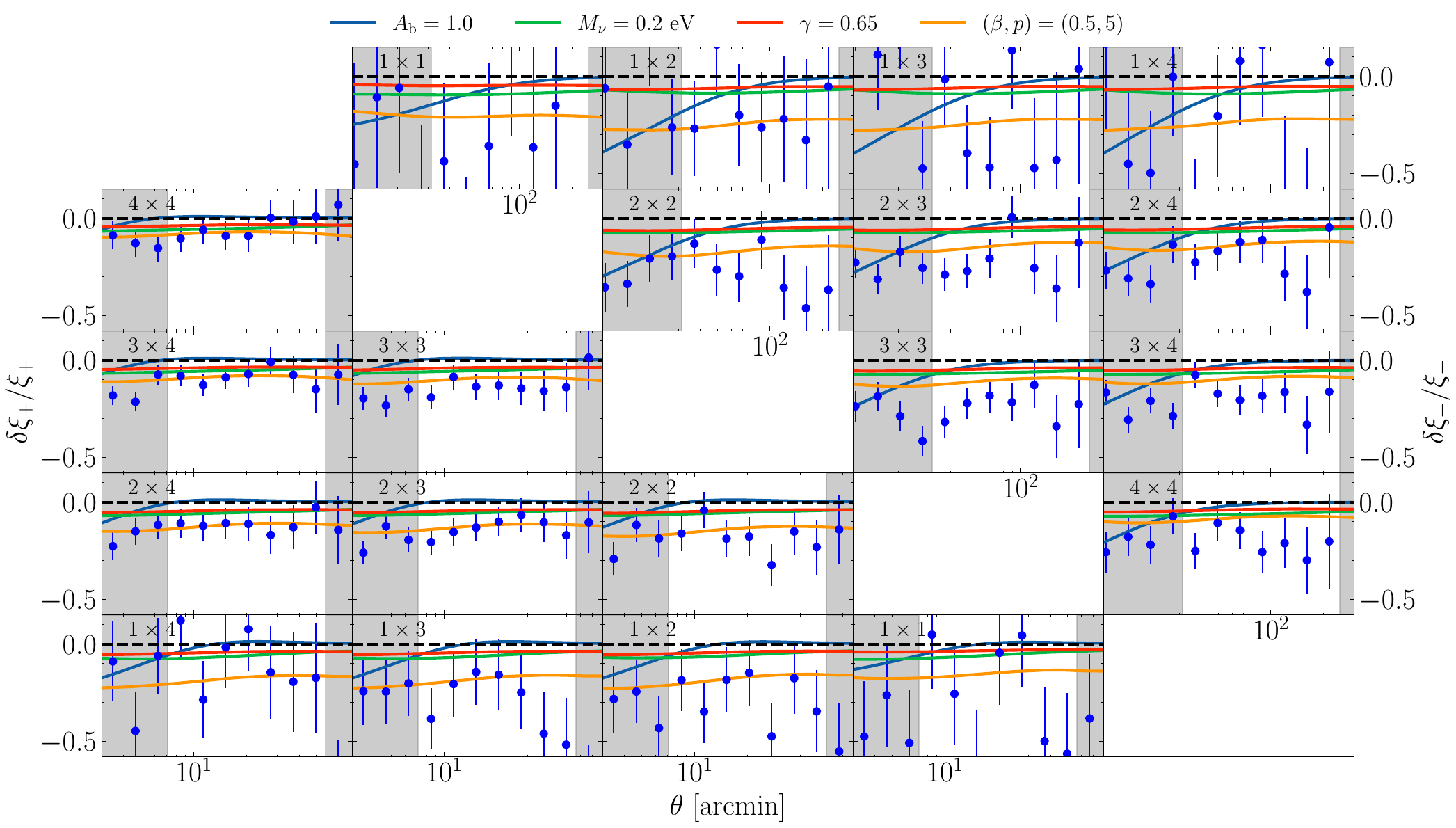}
    \caption{Fractional change in the cosmic shear two-point correlation functions 
    (2PCFs), $\xi_{\pm}(\theta)$, relative to those for the fiducial $\Lambda$CDM model.
     The lower-left diagonal panels are for $\xi_+(\theta)$, while 
    the upper-right diagonal panels are for $\xi_-(\theta)$. Different panels show the auto- and cross-2PCFs for galaxies in two tomographic redshift bins; for instance, $\xi_{\pm}$ for ``$3\times 4$'' are the 2PCFs of source galaxies 
    in the 3rd- and 4th- redshift bins. 
    {As in Figs.~\ref{fig:Pkratio} and \ref{fig:Pzratio}, we consider
    the models with baryonic effects, 
    {the massive neutrinos model,}
    the $\gamma$ growth index model, and the DETG model as extended models.
    We chose model parameter values for each model that 
        lead to a suppression of the cosmic shear 2PCF amplitudes.
         {For comparison, 
    the data points with error bars denote the fractional difference between the measurements from the HSC Year~3 data and the fiducial $\Lambda$CDM model. }
        The unshaded region on the $x$-axis in each panel denotes the range of angular scales used for parameter inference.}
    }
    \label{fig:Effects_on_2PCFs}
\end{figure*}

{In this section, we}
{study how the modifications in structure growth discussed above impact the cosmic shear signal.}

With the flat-sky approximation, the cosmic shear two-point correlation functions (2PCFs) can be expressed 
in terms of the $E$- and $B$-mode angular power spectra $C^{E/B}(\ell)$ via the Hankel transform:
\begin{align}
\label{eq:model_hankel}
\xi^{ij}_{+/-}(\theta) = 
\int\!\frac{\ell \mathrm{d}\ell}{2\pi} \,
        \left[ C^{E;ij}(\ell) \pm C^{B;ij}(\ell) \right]J_{0/4}(\theta \ell) ,
\end{align}
where $J_{0/4}$ are the $0$-th/$4$th-order Bessel functions of the first kind and the transformation of $\xi_+ (\xi_-)$
uses $J_0 (J_4)$.
In our analysis, we use \textsc{FFTLog}
\citep{fftlog2020} to perform 
the Hankel transform.
The superscripts ``$ij$'' denote tomographic bins; e.g. ``$ij$'' means the 2PCF or power spectrum obtained using 
source galaxies in the $i$-th and $j$-th tomographic redshift bins. 

The $E$- {and $B$-}mode angular power spectra in Eq.~(\ref{eq:model_hankel}) are given as
\begin{align}
C^{E;ij}(\ell)&=\int^{\chi_H}_0\!\mathrm{d}\chi\frac{q_i(\chi)q_j(\chi)}{\chi^2}P_{\rm m}\!\!\left(k=\frac{\ell+1/2}{\chi};z(\chi)\right) 
\label{eq:cl_EE}
\end{align}
and
\begin{align}
C^{B;ij}(\ell)&= 0,
\label{eq:cl_BB}
\end{align}
where $\chi(z)$ is the radial comoving distance up to redshift $z$, $\chi_H$ is the distance to the horizon, and 
$q_i(\chi)$ is the lensing efficiency function for source galaxies in the $i$-th redshift bin, defined as
\begin{align}
q_i(\chi)=\frac{3}{2}\Omega_{\rm m}H_0^2\frac{\chi}{a(\chi)}\int_\chi^{\chi_H}\!\mathrm{d}\chi'~n_i(\chi')\frac{\chi'-\chi}{\chi'},
\end{align}
where 
$H_0$ is the present-day Hubble constant ($H_0=100h~{\rm km}~{\rm s}^{-1}~{\rm Mpc}^{-1}$), 
and $n_i(\chi)$ is the normalized redshift distribution of galaxies in the $i$-th source redshift bin. 

{Following \citet{HSC3_cosmicShearReal},
we adopt the fiducial scale cuts;
{we use the data vector of $\xi_+$ in the range}
$\theta \in [7.1, 56.6]$~arcmin and the data of $\xi_-$ in the range 
 $\theta \in [31.2, 247.8]$~arcmin, respectively. 
The wavenumbers of the matter power spectrum 
contributing to $\xi_{\pm}$ at 
these scales are $k \sim [10^{-2}, 1]~h~{\rm Mpc}^{-1}$ (see Fig. 1 of \citet{Terasawa24}).}

{Fig.~\ref{fig:Effects_on_2PCFs} illustrates how the extended models incorporating
the baryonic effect, the varying neutrino mass, 
the $\gamma$ growth index, and the DETG model modify the 
cosmic shear 2PCFs, compared to the predictions for the {fiducial $\Lambda$CDM model, which is consistent with the constraints from {\it Planck}-2018 CMB analysis~\citep{cmb_Planck2018_Cosmology}.}
The varying neutrino mass model and the $\gamma$ growth index model 
cause a suppression in the 2PCF amplitudes over the range of scales considered here.
Both effects exhibit weak scale dependence. 
The DETG model with rapid redshift evolution, specified by $p=5$, modifies the 2PCFs across 
all scales, exhibiting stronger redshift evolution but only weak scale dependence in each tomographic redshift bin.
On the other hand, even the maximum baryonic effect, specified by $A_{\rm b}=1$, 
leaves $\xi_+$ nearly unchanged over the scales in the unshaded region, which are used for cosmological analysis. In other words, the scale cuts are chosen to 
effectively eliminate the scales severely affected by the baryonic effect~\citep{Terasawa24}.
Summarizing the results in Fig.~\ref{fig:Effects_on_2PCFs},
 it is important to note that the different effects exhibit distinct scale and redshift dependencies, which help to distinguish these effects in the cosmological inference from cosmic shear 2PCFs.}

\section{Analysis}
\label{sec:analysis}

\subsection{Data}

\begin{table}
\caption{
Model parameters and priors used in our cosmological parameter
inference. The label ${\mathcal U}(a,b)$ denotes a uniform flat prior
between $a$ and $b$, and ${\mathcal N}(\mu, \sigma)$ denotes a normal
distribution with mean $\mu$ and width $\sigma$.
Our analysis uses six cosmological parameters and additional one or six parameter(s).
We fix the nuisance parameters associated with the cosmic shear 2PCFs to be the MAP values in \citet{HSC3_cosmicShearReal} (see text for details).
}
\label{tab:parameters}
\setlength{\tabcolsep}{20pt}
\begin{center}
\begin{tabular}{ll}  \hline\hline
Parameter & Prior \\ \hline
\multicolumn{2}{l}{\hspace{-1em}\bf Cosmological parameters}
\\
$\ln 10^{10}A_\mathrm{s}$   & ${\cal U}(1.61, 4.0)$\\
$n_\mathrm{s}$                      & ${\cal U}(0.9, 1.1)$\\
$h$                               & ${\cal U}(0.2, 1.0)$\\
$\omega_\mathrm{b}$                 & ${\cal U}(0.005, 0.1)$\\
$\omega_\mathrm{c}$                 & ${\cal U}(0.001, 0.99)$\\
$\tau$                    & ${\cal N}(0.0506, 0.0086)$ \\
\hline \hline
\multicolumn{2}{l}{\hspace{-1em}\bf Additional parameter(s) of extended model} \\
\hline
\multicolumn{2}{l}{\hspace{-1em}\bf massive neutrinos} \\
$M_{\nu}$                 & ${\cal U}(0.0, 5.0)$\\
\hline
\multicolumn{2}{l}{\hspace{-1em}\bf \textsc{HMCode16} baryonic feedback} \\
$A_{\rm b}$                 & ${\cal U}(1.0, 4.0)$\\
\hline
\multicolumn{2}{l}{\hspace{-1em}\bf \textsc{HMCode20} baryonic feedback} \\
$B_0$                               & ${\cal U}(2.35, 4.00)$ \\
$B_z$                            & ${\cal U}(-0.149, -0.039)$ \\
$f_{*,0}$                               & ${\cal U}(0, 0.0265)$ \\
$f_{*,z}$                            & ${\cal U}(0.39, 0.46)$ \\
$\log_{10}(M_{{\rm b}, 0} / h^{-1} M_{\odot})$                     & ${\cal U}(0, 17.85)$ \\
$M_{{\rm b}, z}$                     & ${\cal U}(-0.16, 0.57)$ \\
\hline
\multicolumn{2}{l}{\hspace{-1em}\bf growth index} \\
$\gamma$                 & ${\cal U}(0.0, 2.0)$\\
\hline
\multicolumn{2}{l}{\hspace{-1em}\bf DETG} \\
$\beta$                 & ${\cal U}(-1.0, 1.0)$\\
\hline
\hline
\end{tabular}
\end{center}
\end{table}

To investigate the possibility of mitigating the $S_8$ tension with the models mentioned above, we jointly analyze the cosmological datasets:
the HSC-Y3 cosmic shear 2PCFs~\citep{HSC3_cosmicShearReal},
the ACT DR6 CMB lensing analysis~\citep{ACTDR6_Qu, ACTDR6_Madhavacheril},
the \textit
{Planck}-2018 CMB analysis~\citep{cmb_Planck2018_Cosmology}, and the DESI Y1 BAO analysis~\citep{DESI_BAO}. We summarize the observations as
follows. 

\begin{enumerate}
    \item \textbf{HSC-Y3}: The HSC-Y3 {weak lensing measurements
    were based on the catalog of 25~million galaxies over about 416~deg$^2$}
        (with $n_\text{eff}$$\sim$$15~\mathrm{arcmin}^{-2}$) 
        \citep{HSC3_catalog_Li2021}. 
        {In this paper, we use the data vector containing}
        the cosmic shear 2PCFs measured from the HSC-Y3 data in \citet{HSC3_cosmicShearReal}.
        We employ the fiducial scale-cut in \citet{HSC3_cosmicShearReal} ($\theta \in [7.1, 56.6]$ arcmin for $\xi_+$ and $\theta \in [31.2, 247.8]$ arcmin for $\xi_-$), which corresponds to $k \sim [10^{-2}, 1]~{\rm Mpc}^{-1}$~\citep{Terasawa24}. 
    \item \textbf{ACT DR6:} 
    We use the CMB lensing power spectrum measured from ACT DR6 \citep{ACTDR6_Qu, ACTDR6_Madhavacheril}.
     {We {use the data vector defined within}
     the baseline multipole range in \cite{ACTDR6_Qu, ACTDR6_Madhavacheril}, $40 < L
     < 763$}, which corresponds to $k \sim [4\times 10^{-3}, 0.2]~{\rm Mpc}^{-1}$.
    \item \textbf{Planck-2018:} {We use} 
    the third data release (PR3) from the
        \textit{Planck} 
        experiment
        \citep{cmb_Planck2018_Cosmology}. 
        We {use the data vector containing}
        the primary $TT$, $TE$, and $EE$ {power spectra in the multipole range}
        $30< \ell < 2508$, which corresponds to $k \sim [2\times 10^{-3}, 0.2]~{\rm Mpc}^{-1}$.
         We do not incorporate the low-$\ell$ ($2 <
        \ell < 30$) likelihood. Instead, we {employ a Gaussian prior}
        on the optical depth $\tau$ inferred from the
        low-$\ell$ $EE$ data: ${\cal N}(0.0506, 0.0086)$.
    \item \textbf{DESI Y1 BAO:} We include {the information of}
    baryon acoustic
        oscillation (BAO) measurements from the DESI Year~1 galaxy sample \citep{DESI_BAO}. We
        use {the BAO information for all galaxy samples,}
       ``BGS", ``LRG1", ``LRG2", ``LRG3+ELG1", ``ELG2", ``QSO", and ``Lya QSO".
        {The BAO information help constrain 
        the cosmological parameters of $\Lambda$CDM model, such as $\Omega_{\rm m}$, when
         combined with the {\it Planck} CMB information.} 
\end{enumerate}

\subsection{Parameter inference}

Our analysis uses a set of parameters and priors summarized in Table~\ref{tab:parameters}.
The parameters include six cosmological parameters, denoted by $\mathbb{C}=\left\{\ln 10^{10} A_s, n_s,
h, \omega_{\rm b}, \omega_{\rm c}, \tau \right\}$,
for flat $\Lambda$CDM cosmologies.
$h$ is the Hubble parameter, and 
$\omega_{\rm c}(\equiv \Omega_{\rm c}h^2)$,
$\omega_{\rm b}(\equiv \Omega_{\rm b}h^2)$,
and $\omega_\nu$ 
are the physical density parameters of CDM, baryon and massive neutrinos, respectively. 
{The physical density parameter of non-relativistic matter is given as
$\omega_{\rm m}=\omega_{\rm c}+\omega_{\rm b}+\omega_{\rm \nu}$.
The matter density parameter is given as $\Omega_{\rm m}=\omega_{\rm m}/h^2$.
Except for cases where the neutrino mass is treated as a free parameter, we adopt}
a fixed total neutrino mass of $M_\nu=0.06~{\rm eV}$.
{$A_s$ and $n_s$ are the amplitude and spectral index parameters of the primordial power spectrum
of the curvature perturbations. $\tau$ is the optical depth parameter.}

{The cosmic shear 2PCFs are affected by systematic effects such as baryonic feedback, 
intrinsic alignments, and photometric redshift errors. In this paper, 
we fix the parameters to account for the systematic effects to the values 
at the maximum a posteriori (MAP) $\Lambda$CDM model found in 
\citet{HSC3_cosmicShearReal}, except in cases that 
the baryonic effect parameters are treated as free parameters.}

{For extended models, we introduce additional parameters specific to each model.
In the model with a varying neutrino mass, we treat the neutrino mass as a free parameter in the parameter inference and adopt a flat prior on the neutrino mass, given in Table~\ref{tab:parameters}.}

{For the models to include the baryonic effects, we use either of the \textsc{HMCode16} or \textsc{HMCode20} described in Sections~\ref{subsubsec:HMCode16} and \ref{subsubsec:HMCode20}.
We introduce the parameters and priors given in Table~\ref{tab:parameters}.}

{For the modified growth factor model, we introduce $\gamma$ for the $\gamma$ growth index model
or $\beta$ for the DETG model as a free parameter, respectively, and adopt priors, as given in Table~\ref{tab:parameters}.
Note that, for the DETG model, the other parameter $p$ is assumed to be a specific value, 
and $\beta$ is estimated for each case.
We incorporate the effects of the modified growth factor when computing the theoretical predictions for the CMB lensing signals in both the {\it Planck} and ACT datasets.}

We use \textsc{CAMB}~\citep{CAMB2000} to calculate the primary CMB power spectra and the CMB lensing power spectrum for the $\Lambda$CDM model and the extended models within the $\Lambda$CDM model.
{For the $\gamma$ growth index model and 
the DETG model, we use the modified version of \textsc{CAMB} to compute the linear power spectrum 
at a given redshift, as described above. Then, we apply a nonlinear mapping to the input linear power spectrum to compute the nonlinear power spectrum (again see above).}

We use the CMB and CMB lensing likelihood where the associated nuisance parameters are already marginalized over. 
We assume the {different datasets}
are independent and ignore the cross-covariance between  
them.

We use \textsc{Multinest}~\citep{multinest_Feroz2009} to sample the posterior distribution.
Throughout this paper, we report the 1D marginalized {\em mode} and its asymmetric 68\% credible intervals,
together with the MAP estimated as the maximum of the posterior in the chain
\citep[see Eq.~1 in][]{HSC1_2x2pt_Miyatake2022}. For the 2D marginalized posterior, 
we report the mode and the 68\% and 95\% credible {region}. 
We use \textsc{GetDist} \citep{GetDist2019} for the plotting.

To compare the different models {of growth suppression with}
the $\Lambda$CDM model, we use goodness-of-fit $\chi^2 = -2\ln \mathcal{L}^{\rm MAP}$.
In practice we compare the difference in the $\chi^2$ values divided by the number of the additional parameters $\Delta N_p$, 
namely $\Delta \chi^2/\Delta N_p$, between the extended model and the $\Lambda$CDM model.
Since $\Delta \chi^2$ follows the $\chi^2$ distribution with $\Delta N_p$ degrees of freedom, its typical value is $\Delta \chi^2/\Delta N_p \simeq 1$. $|\Delta \chi^2/\Delta N_p| > 1 (< 1)$ {indicates a better (worse) fit when accounting for} the number of the additional parameters, and 
{suggests a preference (disfavor)}
for the extended model {compared to the $\Lambda$CDM model}. 
We also use the Bayes factor $R = {\mathcal{Z}/\mathcal{Z}_{\Lambda{\rm CDM}}}$,
where $\mathcal{Z}$ is the Bayesian evidence obtained from \textsc{Multinest} sampler and $\mathcal{Z}_{\Lambda{\rm CDM}}$ denotes the Bayesian evidence for the $\Lambda$CDM model.

\section{Results}
\label{sec:results}

\begin{table*}
     \centering
     \caption{{Performance metrics of extended models, compared to the $\Lambda$CDM model.}
     {The column ``best-fit" denotes the value of the extended model parameter at MAP for the single-parameter extensions. 
     The column ``mode and $68\%$ CL" denotes the mode value and 
    the 68\% credible interval for the parameter,
    for each model. When the mode is located at the lower bound of the prior, we only show the upper limit of 
    the 68\% credible interval. 
     The column ``$\Delta\chi^2/\Delta N_p$" denotes $\Delta\chi^2$ divided by the number of the additional parameters $\Delta N_p$.
     The column ``$\ln R$" denotes the logarithmic Bayes factor.}
     }
     \begin{tabular}{l|c|c|c|c} \hline \hline
        model  & best-fit & mode and $68\%$ CL & $\Delta \chi^2/\Delta N_p$ & $\ln R$ \\ \hline
     \textsc{HMCode16} baryon feedback & $A_{\rm b} = 1.03$ & $A_{\rm b} < 2.11$  &$1.6$ & $0.2$ \\ %\hline
     \textsc{HMCode20} baryon feedback & -- & -- & $0.75$ & $0.3$ \\ %\hline
    Massive neutrinos & $M_{\nu} = 0.0026~$eV& $M_{\nu} < 0.033 $  & $2.1$ & $-3.3$  \\ %\hline
      $\gamma $ growth index & $\gamma = 0.65$ & $\gamma = 0.69^{+0.135}_{-0.108}$  & $1.4$ & $-1.0$
  \\ %\hline
      DETG ($p=1$) & $\beta = 0.09$&$\beta = 0.074^{+0.067}_{-0.078}$  & $2.0$ & $-1.8$ \\ %\hline
     DETG ($p=2$) & $\beta = 0.18$&$\beta = 0.152^{+0.111}_{-0.094}$  & $3.4$ & $-0.4$ \\ %\hline
     DETG ($p=3$) & $\beta = 0.29$&$\beta = 0.251^{+0.151}_{-0.123}$  & $5.7$ & $0.3$ \\ %\hline
     DETG ($p=4$) & $\beta = 0.33$&$\beta = 0.367^{+0.168}_{-0.176}$  & $4.1$ & $0.7$ \\ %\hline
     DETG ($p=5$) & $\beta = 0.53$&$\beta = 0.443^{+0.227}_{-0.203}$  & $5.1$ & $1.0$ \\ %\hline
     DETG ($p=6$) & $\beta = 0.69$&$\beta = 0.578^{+0.260}_{-0.223}$  & $4.8$ & $1.1$ \\ \hline
      \hline
     \end{tabular}
     \label{tab:goodness-of-fit}
 \end{table*}
 
\begin{figure}
\includegraphics[width=\columnwidth]{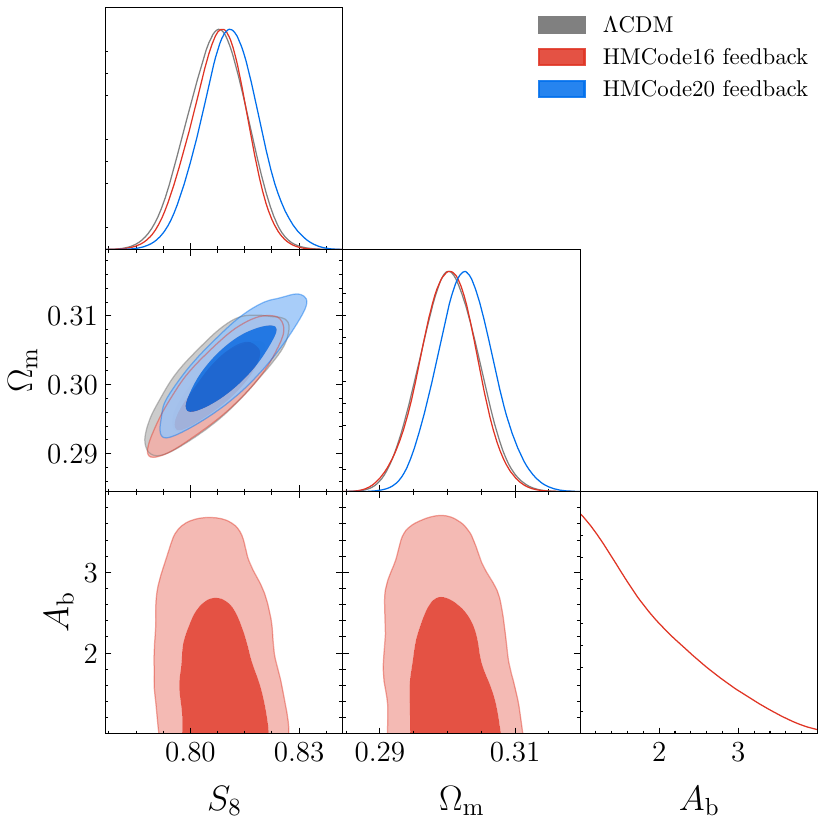}
    \caption{
    The 1D and 2D posteriors obtained from the parameter inference 
    of the joint datasets of the HSC-Y3 cosmic shear 2PCFs, the {\it Planck} CMB spectra, the ACT DR6 CMB lensing spectrum, and the DESI-Y1 BAO data. The red and blue contours show the results for the $\Lambda$CDM model including the baryonic effects modeled by \textsc{HMCode16} (Section~\ref{subsubsec:HMCode16}) 
    or by \textsc{HMCode20} (Section~\ref{subsubsec:HMCode20}).
    $A_{\rm b}$ is the parameter of the \textsc{HMCode16}, where $A_{\rm b}=3.13$ corresponds to the DM-only prediction, and the lower values correspond to the stronger feedback effects.
    }
    \label{fig:2D_baryon}
\end{figure}

\begin{figure}
\includegraphics[width=\columnwidth]{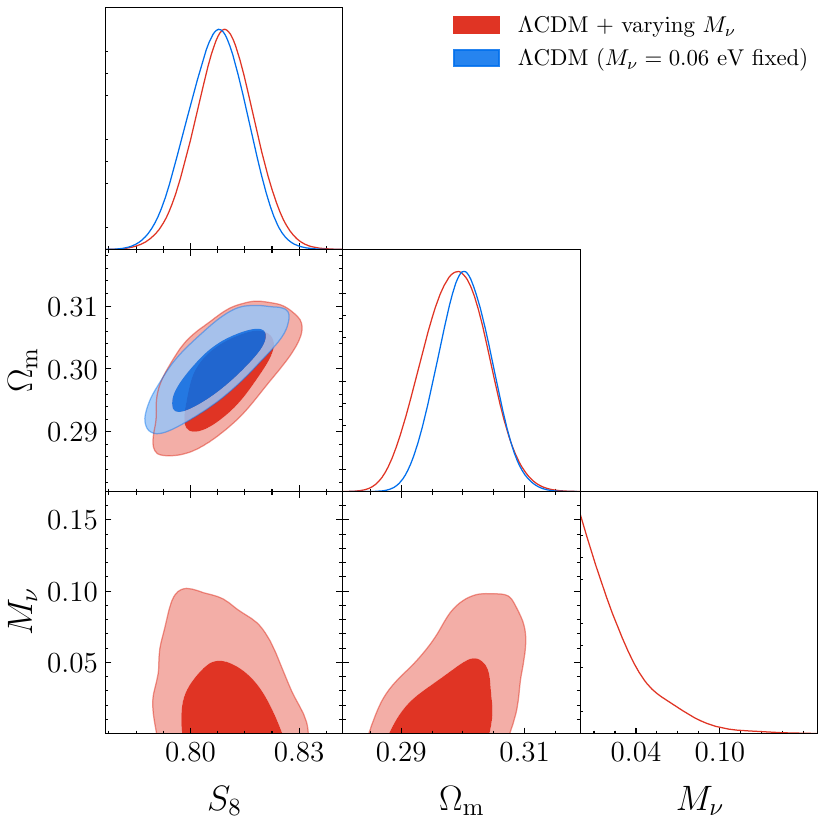}
    \caption{The 1D and 2D posteriors obtained from the parameter inference 
    of the $\Lambda$CDM model (blue contours) and the $\Lambda$CDM $+ M_{\nu}$ model (red contours).}
    \label{fig:2D_mnu}
\end{figure}

\begin{figure}
\includegraphics[width=\columnwidth]{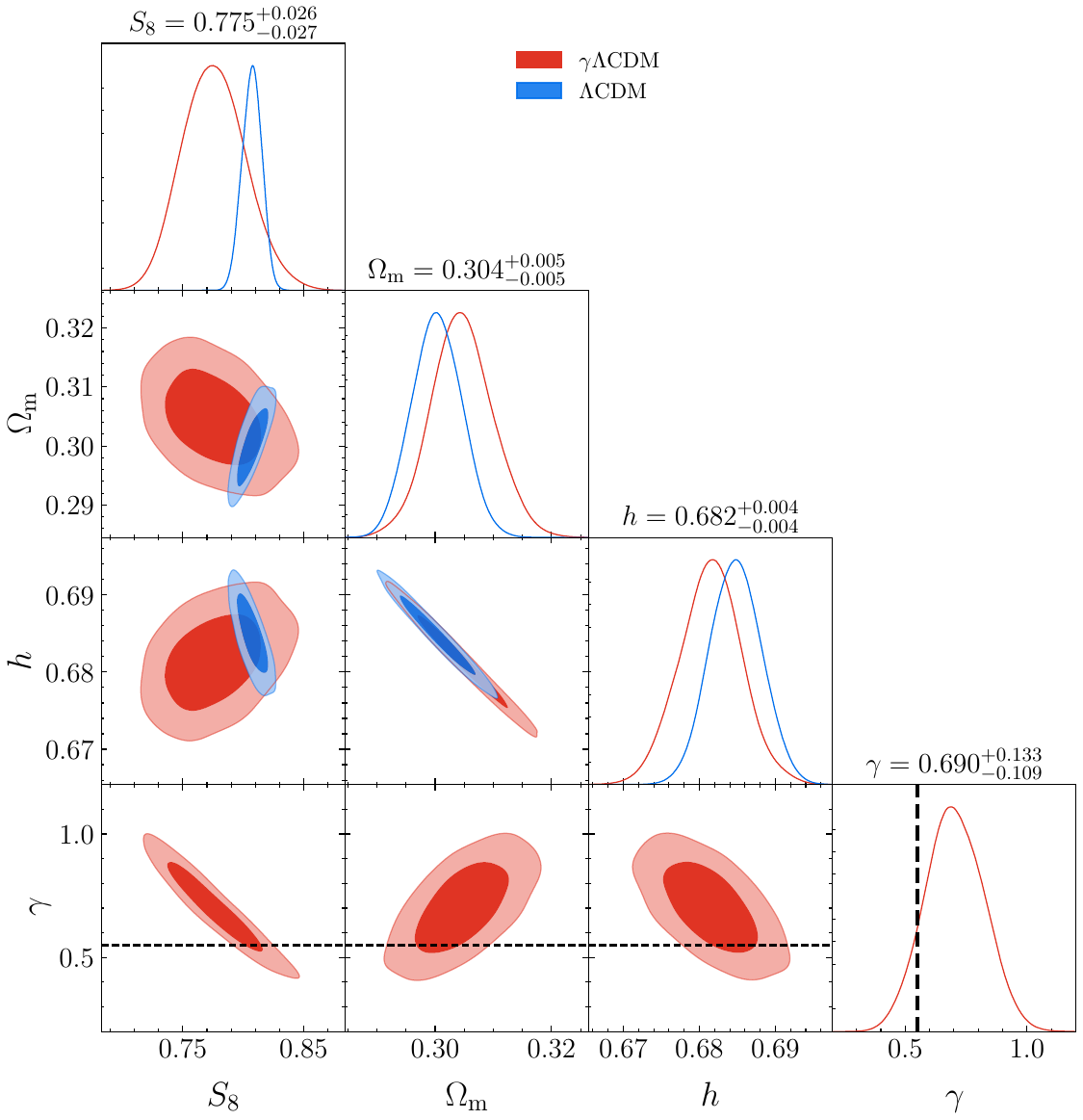}
    \caption{The 1D and 2D posteriors obtained from the parameter inference 
    of the $\Lambda$CDM model (blue contours) and the $\gamma$ growth index model
(red contours, 
    denoted as ``$\gamma\Lambda$CDM"). {The dashed lines in the bottom low panels denote the $\Lambda$CDM limit, $\gamma=0.55$.}
    The numbers above each panel 
    of the 1D posterior denote the mode value and 
    the 68\% credible interval for the parameter,
    for the ``$\gamma\Lambda$CDM'' result.
    }
    \label{fig:2D_gamma}
\end{figure}

\begin{figure*}
\includegraphics[width=2\columnwidth]{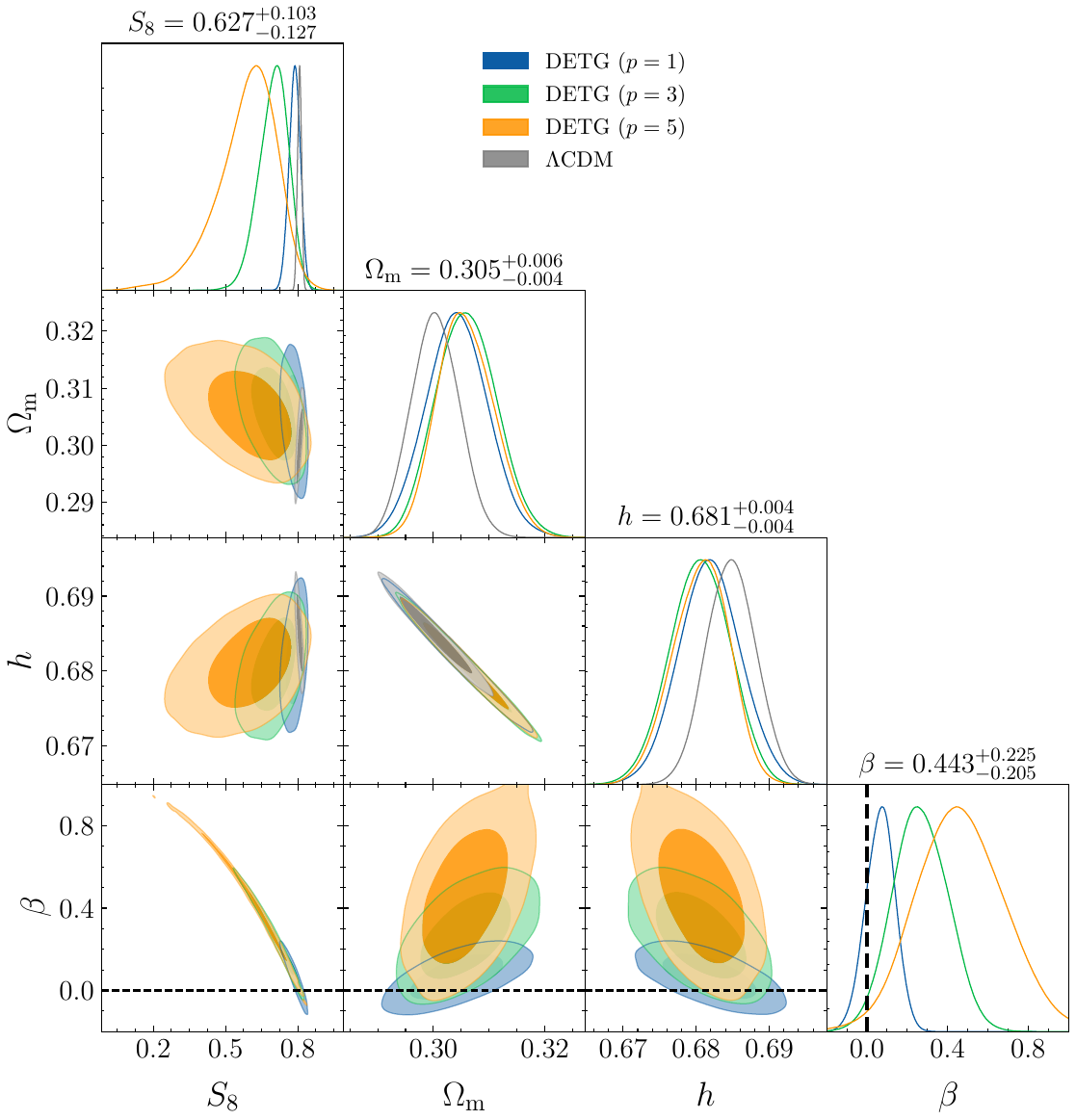}
    \caption{The 1D and 2D posteriors obtained from the parameter inference 
    of the $\Lambda$CDM model (gray contours) and the DETG model.
    {The dashed lines in the bottom row panels denote the $\Lambda$CDM limit, $\beta=0$.}
    The numbers above each panel 
    of the 1D posterior denote the mode value and 
    the 68\% credible interval for the parameter,
    for the ``DETG $(p=5)$'' result.
    }
    \label{fig:2D_DETG}
\end{figure*}

\subsection{Baryonic effect}
\label{sec:results_baryon}

{In this subsection, we study the baryonic effect model as an extended model. The red contours in}
Fig.~\ref{fig:2D_baryon} show the posterior distributions obtained from the analysis with varying the baryonic feedback parameter $A_{\rm b}$ {using \textsc{HMCode16}} (see Section~\ref{subsubsec:HMCode16}). 
$A_{\rm b}$ {exhibits only a weak}
correlation with $S_8$ in the 2D posterior, and {the} inclusion of $A_{\rm b}$ has little impact on the posterior distributions of the other $\Lambda$CDM parameters. 
The improvement in the goodness of fit is $\Delta \chi^2 = 1.6$ as given in Table~\ref{tab:goodness-of-fit}, 
which is a typical value for 
{the inclusion of one additional parameter.}
The Bayes factor compared to the $\Lambda$CDM model is $\ln R = 0.2$, 
{indicating a marginal preference for the extended model. 
According to Jeffreys' scale~\citep{Jeffreys}, this value is {considered ``barely worth mentioning''.}

{The blue contours in}
Fig.~\ref{fig:2D_baryon}
show the posteriors in the ($S_8$, $\Om$) plane, obtained from the analysis using
the HMCode20 model which includes six additional parameters (Section~\ref{subsubsec:HMCode20}). 
{The inferred value of $S_8$ is slightly}
larger than that obtained with the HMCode16 model, {indicating that a stronger feedback model is acceptable.} However, {the $\chi^2$ value for the best-fit value improves}
 by only 4.5 (see Table~\ref{tab:goodness-of-fit}) compared to that for the best-fit $\Lambda$CDM model, which is not {statistically} significant 
 {given}
 the {inclusion of six additional} parameters.
The Bayes factor is $\ln R = 0.3$, which is similar to that of HMCode16 model and 
{considered}
``barely worth mentioning". Hence we conclude that the baryonic effect {models}
are not favored over the $\Lambda$CDM model.

\subsection{Massive neutrinos}
\label{sec:results_neutrinos}

In Fig.~\ref{fig:2D_mnu} {we present the results obtained by varying the total neutrino mass, $M_\nu$, in the inference, under the framework of the $\Lambda$CDM model and the GR gravity. The figure shows that allowing the neutrino mass to vary does not mitigate the $S_8$ tension; 
the cosmological parameters remain nearly unchanged, compared to the $\Lambda$CDM model with a fixed neutrino 
mass of $M_\nu=0.06$~eV. As given in Table~\ref{tab:goodness-of-fit}, the model with a varying neutrino mass does not lead to a significant improvement in the $\chi^2$ value.}
The Bayes factor is as low as $\ln R = -3.3$, indicating a strong preference for $\Lambda$CDM model with $M_\nu=0.06$~eV.
{Therefore, we conclude that the scale and redshift dependence of the suppression in structure growth caused by massive neutrinos is not favored by the data.}

\subsection{$\gamma$ growth index model}
\label{sec:results_gamma}

{In this subsection, we study the $\gamma$ growth index model. The red contours in}
Fig.~\ref{fig:2D_gamma} show the posterior distributions obtained from the analysis {including the $\gamma$ parameter as an additional 
parameter in the $\gamma$ growth index model (Section~\ref{ssub:growth_index}), compared to the blue contours obtained from the analysis based on the $\Lambda$CDM model. 
The joint dataset of {\it Planck} CMB, ACT DR6 CMB lensing, and HSC-Y3 cosmic shear data favors 
a $\gamma$ value higher than the GR prediction of $\gamma\simeq 0.55$, {although 
the posterior distribution of $\gamma$ still includes $\gamma=0.55$.}
The higher $\gamma$ model leads to a suppression in structure growth at lower redshifts relevant to the cosmic shear data (see Figs.~\ref{fig:Pkratio} and \ref{fig:Effects_on_2PCFs}), thereby leading to a lower value of $S_8$ for the CMB normalization of the linear power spectrum.
{The $\gamma$ growth index model marginally improves the $\chi^2$ value for the best-fit model compared to the $\Lambda$CDM model, with 
$\Delta\chi^2=1.4$, as given in Table~\ref{tab:goodness-of-fit}. The Bayes factor is as low as $\ln R=-1.0$, which favors the $\Lambda$CDM model
over the $\gamma$ model.
}

\subsection{DETG}
\label{sec:results_DETG}

{In Fig.~\ref{fig:2D_DETG}, we show the results for the DETG model. The colored lines 
show}
the posterior distributions obtained from the analysis {using the DETG model (Section~\ref{ssub:detg}), compared to those 
for the $\Lambda$CDM model. While the DETG model is given by two parameters ($\beta, p$), Fig.~\ref{fig:2D_DETG} shows the results for a fixed
value of $p=1, 3$ or 5, respectively, with varying $\beta$ in the inference. 
Note that the model with $\beta=0$ corresponds to 
the $\Lambda$CDM model, or equivalently, the GR model.
The figure 
shows that} the posteriors favor {the model with} $\beta>0$, indicating that the data prefer a suppression of structure growth in the late universe.
{As summarized in Table~\ref{tab:goodness-of-fit}, the improvement in the $\chi^2$ value for the best-fit model
is substantial with $\Delta \chi^2/\Delta N_p\gtrsim 4$ 
for the DETG model with $p=3,4, 5$ or $6$, which corresponds to a significance level of more than 
a $2\sigma$ level under a naive Gaussian-like consideration. However, the Bayes factor is  not particularly large:  
the DETG model with $p=3, 4, 5$ or $6$ has $\ln R>0$, but are categorized as ``barely worth mentioning''.
}

{As seen in the Fig.~\ref{fig:Pzratio}, the redshift dependence of the growth suppression of the $\gamma$ index model is close to that of the DETG model with $p=1$. That explains why these two models have similar $\Delta \chi^2/\Delta N_p$ and $\ln R$.
}

\section{Conclusion}
\label{sec:conclusion}

{In this paper, we have considered extended models within or beyond the $\Lambda$CDM framework to investigate whether such models can explain both early- and late-time cosmological datasets, 
the {\it Planck} CMB data, the ACT DR6 CMB lensing data and the Subaru HSC-Y3 cosmic shear data, 
motivated by the $S_8$ tension.}
We considered the extended models incorporating 
the baryonic feedback effects, the neutrino mass uncertainty, the $\gamma$ growth index model, and the Dark Energy Tracking Growth (DETG) model.

Our findings are summarized as follows:
\begin{itemize}
    \item 
We showed that the modified growth factor models, especially those featuring more rapid growth suppression at lower redshifts, provide an improved fit to the combined datasets. 
For the DETG model with $p=3,4,5$ or $6$, 
We found a potential deviation from the GR prediction at the
$\sim 2\sigma$ level, 
with $\beta > 0$. 
\item {We showed that neither the model incorporating parameterized baryonic feedback effects nor the one treating neutrino mass as a free parameter is preferred over}
the $\Lambda$CDM model. 
\end{itemize}
An interesting implication of our results is that models featuring a rapidly evolving modification to the growth factor toward the present epoch are favored (see Fig.~\ref{fig:Pzratio}), such as 
the DETG model with $p=5$, compared to the 
$\gamma$ growth index model or the DETG model with $p=1$ or 2. 
Such rapid evolution toward the present epoch 
{is consistent with the results from the tomographic cross-correlation of DESI luminous red galaxies and CMB lensing~\citep{2022JCAP...02..007W, 2024arXiv240704607S} and the combined analysis of cosmic shear, galaxy clustering, and CMB lensing~\citep{2021JCAP...10..030G}.
The rapid evolution at low redshifts is seen not only in the structure growth, but also in the background expansion, explained by time-varying dark energy~\citep[e.g., the recent DESI BAO measurements][]{2025arXiv250314738D}.
}
These implications might be physically related, but could also be mere coincidence. More data are needed to establish the statistical significance of these findings, but we believe they are worth mentioning as they are highly intriguing.

\twocolumngrid

\acknowledgments
We would like to thank Minh Nguyen and Meng-Xiang Lin for useful discussion. 
This work was supported in part by JSPS KAKENHI Grant Numbers 20H05850,
20H05855,
and 23KJ0747,
and
by World Premier International Research Center Initiative (WPI Initiative), MEXT, Japan.
SS is supported by the JSPS Overseas Research Fellowships.

\bibliography{refs-short}

\end{document}